\documentclass[graybox]{svmult}

\usepackage{type1cm}        
\usepackage{makeidx}         
\usepackage{graphicx}        
\usepackage{multicol}        
\usepackage[bottom]{footmisc}

\usepackage{newtxtext}       %
\usepackage{newtxmath}       

\usepackage{amsmath}
\usepackage{amsfonts}
\usepackage{bm}
\usepackage{hyperref}
\hypersetup{
	colorlinks   = true,     
	urlcolor     = blue, 
	linkcolor    = blue, 
	citecolor  	 = blue  
}
\usepackage[caption=false,position=top]{subfig}
\usepackage{dsfont}

\def\Bext{B_\mathrm{ext}}
\def\Bov{\bm{B}_\mathrm{ov}}
\def\Beff{B_\mathrm{eff}}

\def\ex{\bm{e}_x}
\def\ey{\bm{e}_y}
\def\ez{\bm{e}_z}
\def\S{\bm{S}}
\def\J{\bm{J}}

\def\Ntr{N_\mathrm{tr}}
\def\Neff{N_\mathrm{eff}}
\def\TR{T_\mathrm{R}}
\def\Tnstar{T_\mathrm{n}^*}
\def\np{n_\mathrm{p}}
\def\ge{g_\mathrm{e}}
\def\gh{g_\mathrm{h}}
\def\gn{g_\mathrm{n}}
\def\muB{\mu_\mathrm{B}}
\def\mun{\mu_\mathrm{n}}
\def\Slim{S_\mathrm{lim}}
\def\fm{f_\mathrm{m}}
\def\ddt{\frac{\mathrm{d}}{\mathrm{d}t}}
\def\tausg{\tau_\mathrm{s,g}}
\def\taust{\tau_\mathrm{s,t}}
\def\tauseff{\tau_\mathrm{s}^*}
\def\wn{\omega_\mathrm{n,g}}

\makeindex             

\begin{document}

\title*{Simulation of nonequilibrium spin dynamics in quantum dots subjected to periodic laser pulses}
\titlerunning{Nonequilibrium spin dynamics in quantum dots subjected to periodic laser pulses}        

\author{Philipp~Schering, Philipp~W.~Scherer, and G\"otz~S.~Uhrig}

\institute{Philipp Schering \at
	Condensed Matter Theory, TU Dortmund University, 44221 Dortmund, Germany \\
	\email{philipp.schering@tu-dortmund.de}           
	\and
	G\"otz S. Uhrig \at
	Condensed Matter Theory, TU Dortmund University, 44221 Dortmund, Germany \\
	\email{goetz.uhrig@tu-dortmund.de}
}
%
%
\maketitle

\abstract*{Large-scale simulations of the spin dynamics in quantum dots subjected to trains of periodic laser pulses enable us to describe and understand related experiments. By comparing the data for different models to experimental results, we gain an improved understanding of the relevant physical mechanisms. Using sophisticated numerical approaches and an efficient implementation combined with extrapolation arguments, nonequilibrium stationary states are reached for parameter ranges close to the ones in real experiments. With the help of high performance computing, we can tune the experimental parameters to guide future experimental research. Importantly, our simulations reveal the possibility of resonant spin amplification in Faraday geometry, i.e., when a longitudinal magnetic field is applied to the quantum dots.}

\abstract{Large-scale simulations of the spin dynamics in quantum dots subjected to trains of periodic laser pulses enable us to describe and understand related experiments. By comparing the data for different models to experimental results, we gain an improved understanding of the relevant physical mechanisms. Using sophisticated numerical approaches and an efficient implementation combined with extrapolation arguments, nonequilibrium stationary states are reached for parameter ranges close to the ones in real experiments. With the help of high performance computing, we can tune the experimental parameters to guide future experimental research. Importantly, our simulations reveal the possibility of resonant spin amplification in Faraday geometry, i.\,e., when a longitudinal magnetic field is applied to the quantum dots.}

\section{Introduction}
\label{sec:intro}

A localized electronic spin in a semiconductor quantum dot~(QD) is considered a promising
candidate for the realization of quantum bits~\cite{divincenzo98}, which
are at the very basis of any quantum information processing~\cite{niels00}. 
Such an electronic spin in a QD loses its coherence due to 
its interaction with the bath of nuclear spins of the surrounding isotopes in III-V semiconductors.
The number of substantially coupled nuclear spins is very large of the order of $10^4$ to $10^6$~\cite{urba13}.

Considerable effort has been invested in the experimental investigation of the spin dynamics in semiconductor
nanostructures and the possibilities to manipulate it~\cite{urba13,schliemann03,hanson07}.
It is particularly interesting  that ensembles of QDs
can be manipulated as well. They can be made to respond coherently by subjecting them 
to long periodic trains of laser pulses while applying a transverse magnetic field~\cite{greil06a,greil06b,greil07a}. 
Experimentally, the train of periodic pulses is applied for
seconds to minutes which implies up to $10^{10}$ pulses because the 
generic repetition period \,$T_\mathrm{R}$ of the pulses is of the order of $10\,$ns.
Since the period of the electronic Larmor precession is of the order of 10~ps,
the theoretical simulations have to cover 12 orders of magnitude in time for
systems with about $10^5$ spins; this is a tremendous computational challenge
even for high performance computing~(HPC).

It appears that the application of periodic pulses with repetition period \,$T_\mathrm{R}$ synchronizes
the Larmor precessions of the spins in sub-ensembles of QDs, eventually leading to constructive interference of the Larmor precessions before each pulse accompanied by a revival of the spin polarization.
The Overhauser field, i.\,e., the magnetic field applied by all the nuclear spins together
via the hyperfine interaction on the electronic spin, 
changes slightly such that it compensates the fluctuations in the $g$ factor
from dot to dot which otherwise would lead to fast dephasing of the electronic Larmor precessions
of different dots. This phenomenon is called nuclei-induced frequency focusing~\cite{greil07a} and 
it is the effect which we study by quantitative simulations possible
thanks to high performance computing.
Thereby, we pave the way for future experiments exploiting the electronic spin in QDs as a quantum resource.
The long-term goal is to generate coherent states
including the nuclear spin degrees of freedom in single and multiple QDs
by suitable pulse protocols, thereby lifting coherent control to another level.

Recent experiments on spin inertia and polarization recovery in QDs~\cite{zhukov18} 
revealed results which could not be fully explained by the analytic theoretical model~\cite{smirnov18}.
By utilizing the high performance facilities of the HLRS, we can perform improved simulations of 
these experiments~\cite{scher19}, which help us to gain a better understanding of the underlying physics. 
Importantly, we find the emergence of resonant spin amplification in the so called Faraday geometry, i.\,e., when a longitudinal magnetic field is applied to the QDs. This new effect can be revealed experimentally using optimized pulse protocols. Preliminary experimental results confirm its existence.

\section{Nuclei-induced frequency focusing in quantum dots}
\label{sec:NIFF}

We investigate the spin dynamics of an inhomogeneous ensemble of GaAs QDs in a transverse magnetic field (Voigt geometry).
Each QD is singly charged by a localized electron, whose spin couples to the surrounding nuclear 
spins via the hyperfine interaction.
This electronic spin is excited optically by trains of resonant laser pulses with repetition period 
$\TR = 13.2\,$ns, generating negatively charged singlet trion states~(transition energy $\sim1.4\,$eV~\cite{greil06a}).
The trion eventually recombines, inducing some electronic spin polarization into the system due to the spin dynamics in combination with the selection rules.
This polarization dephases on a timescale of nanoseconds due to the random fluctuating 
nuclear spin bath, whose collective hyperfine interaction acts as Overhauser field
on the electronic spin, and due to the spread of the electronic $g$ factors, which differ slightly from dot to dot.
Experimentally, the spin polarization can be probed using weak linearly polarized pulses by measuring the Faraday rotation or ellipticity~\cite{yugov09,glazo12b}.

Upon application of long pulse trains, the spin dynamics reacts in such a way that a revival of 
the spin polarization emerges before the arrival of each next pulse. This effect is known as 
spin mode locking~\cite{greil06b}.
It can be enhanced by the fascinating phenomenon of nuclei-induced frequency focusing~\cite{greil07a}, 
which is one of the central subjects of the present report.
The periodic pumping of the electronic spin indirectly drives the Overhauser field such that 
the effective Larmor frequency of the electronic spin in each QD complies with a 
certain resonance condition. 
Generically, this leads to an enhancement of the mode locked revival amplitude~\cite{greil07a},
but the dependence of this amplitude on the magnetic field strength is complex~\cite{varwig14,jasch17,klein18}.

Here, we report on the recent progress on the simulation of this type of experiment. 
For a more detailed description of the simulations, the results, and the physics, we refer the 
reader to Ref.~\cite{scher20}. 
In order to render the simulation possible, progress in several key areas was required.
First, we enhanced the semiclassical model describing the physical system and 
the optical generation of spin polarization via the excitation of a trion.
The subsequent application of an efficient algorithm to the equations of motion, reducing the 
dimension of the system, is mandatory to deal with large bath sizes~\cite{fauseweh17,scher18}.
In order to be able to make statements for the relevant number of total pulses, i.\,e., 
after which the system is in a nonequilibrium stationary state~(NESS) as in the experiment, a very efficient and 
highly parallel implementation is required to solve the equations of motion.
Even then it is not possible to deal directly with realistic bath sizes of up to $10^6$ 
nuclear spins. We overcome this obstacle using established scaling arguments by which 
we can extrapolate to an infinite bath size.

\subsection{Hyperfine interaction of an electronic spin with a nuclear spin bath}
\label{sec:model_NIFF}

The dominant interaction in a GaAs QD singly charged by electrons is the Fermi contact hyperfine interaction~\cite{urba13}.
In each QD, the the quantum mechanical behavior of the spins is governed by the  Hamiltonian
\begin{equation}
\hat{\mathcal{H}}_\mathrm{hf} = \sum_{k=1}^N A_k \hat{\S} \cdot \hat{\bm I}_k = \hat{\S} \cdot \hat{\bm B}_\mathrm{ov} \,,
\end{equation}
in which the nuclear spins $\hat{\bm  I}_k$ weighted by their hyperfine coupling constant $A_k$ form the so called Overhauser field
\begin{equation}
\hat{\bm B}_\mathrm{ov} = \sum_{k=1}^N A_k \hat{\bm I}_k \,,
\end{equation}
which couples to the central electronic spin $\hat{\S}$.

Solving the full quantum model is extremely restricted in the number of nuclear spins $N$ due to the 
exponentially growing Hilbert space.
We resort to an established semiclassical description of the problem, where the spins are considered as 
classical vectors with random initial conditions~\cite{chen07,stanek14,lindoy20}. 
In this approach, the spin dynamics are governed by the classical equations of motion ($\hbar$ is set to unity)~\cite{scher20}
\begin{subequations}
	\begin{align}
	\ddt \S &= \left( \Bov + \ge \muB \Bext \ex \right) \times \S + \frac{1}{\tau_0} J^z \ez \,, \label{eq:eom_S}
	\\
	\ddt \J &=  \chi \left( B^z_\mathrm{ov} \ez + \frac{1}{\lambda} \Bov^\perp \right) \times \J 
	+ \gh \muB \Bext \ex \times \J - \frac{1}{\tau_0} \J \,, 
	\label{eq:eom_fT} 
	\\
	\ddt \bm I_k &= \left[ A_k \S + \chi A_k \left( J^z \ez + \frac{1}{\lambda} \J^\perp \right) + \gn \mun \Bext \ex \right] 
	\times \bm I_k \,, \label{eq:eom_Ik}
	\end{align}
	\label{eq:eom}%
\end{subequations}
where $\bm B^\perp := B^x \ex + B^y \ey$ and $\J^\perp := J^x \ex + J^y \ey$. 
These equations essentially describe a precession of the classical spins.
The electronic spin~$\S$ precesses around the effective magnetic field 
${\bm{B}_\mathrm{eff} := (\Bov + \ge \muB \Bext \ex)/ (\ge \muB)}$, where $\Bext$ is the strength of the 
external transverse magnetic field, $\ge=0.555$~\cite{greil07a} the $g$ factor of the electronic spin and $\muB$ the Bohr magneton.
The same holds for the trion pseudospin~$\J$, but its hyperfine interaction is weaker by a factor $\chi \approx 0.2$ and 
also anisotropic ($\lambda=5$)~\cite{zhukov18}.
Moreover, the trion decays radiatively on the timescale $\tau_0 = 400\,$ps~\cite{greil06a,greil06b}. 
According to the selection rules, a recombination of the trion component $J^z$ and the ground state $\S$ 
takes place. 
Spin polarization in the ground state is generated when this recombination does not occur
with the exactly same spin quantum number, for instance in an applied transverse magnetic field
inducing Larmor precessions with different frequencies for $\S$ and $\J$.
The nuclear spins $\bm I_k$ also precess around the so called Knight field plus the external magnetic field. 
Due to the larger masses of the nuclei, their gyromagnetic ratio $\gn\mu_\mathrm{n}$ is smaller than $\ge \muB$ by three orders of 
magnitude. Nevertheless, the nuclear Zeeman term has a crucial impact on the nonequilibrium physics since the energy scale of $A_k$ and $\gn \mun \Bext$ can be of similar magnitude for large magnetic fields.
The hyperfine couplings are parameterized according to
\begin{equation}
A_k \propto \exp(-k\gamma) \,, \qquad k \in \{1,\dots,N\} \,,
\label{eq:couplings}%
\end{equation}
which is a realistic choice for flat two-dimensional QDs~\cite{coish04,faribault13a,fauseweh17}.

The ordinary differential equation system has the dimension $3N + 6$, where $N$ is the total number of bath spins. 
Note that $N$ is practically infinite in any solid state system. The number $\Neff$ of effectively coupled spins
within the localization volume of the electronic spin is much smaller, but still very large.
For realistic bath sizes of at least $N=10^4$, the numerical simulation of the desired properties is unfeasible 
even on a HPC system.
We resort to the efficient approach established in Ref.~\cite{fauseweh17}, where sums of bath spins define
auxiliary vectors. It is sufficient to track $\Ntr$ of these auxiliary vectors in a simulation. 
This reduces the dimension to $3\Ntr + 6$, where $\Ntr = \mathcal{O}(75)$ is a truncation parameter, 
while allowing us to treat an infinite spin bath $N \to \infty$ with an effective number $\Neff \approx 2/\gamma$ of sizeably coupled nuclear spins. 
The complexity of the equations does not increase. For details, we refer the interested reader to Ref.~\cite{fauseweh17}, where this approach is established, and to Refs.~\cite{scher18,scher20}, where it is applied to study nonequilibrium spin dynamics in quantum dots.

Due to the large number of nuclear spins, the Overhauser field and the auxiliary fields essentially behave like classical fields.
Furthermore, the large number of contributing spins meets the precondition to use the central limit
theorem to conclude that the Overhauser field and the auxiliary fields are initially normal distributed so that the initial 
auxiliary vectors can be sampled from normal distributions. 
The corresponding mean values and variances are chosen such that they mimic quantum mechanical properties~\cite{fauseweh17}.
The theoretical foundation for this approach is the truncated Wigner approximation~\cite{polkovnikov10}. 
First order quantum fluctuations are taken into account through the correct sampling of the initial conditions for the classical equations of motion.

Moreover, ensembles of QDs are not homogeneous. 
This leads to a slight spread of the $g$ factors of the electronic spin and the trion pseudospin. 
We account for this spread by sampling the $g$ factors from a normal distribution around their 
mean values with appropriate variances $(\Delta g)^2$.

The trion is excited by resonant circularly-polarized laser pulses; we consider so called $\pi$~pulses with helicity $\sigma^-$ here. They have a typical duration of $1.5\,$ps~\cite{greil06a,greil06b,greil07a}, which is 
one order of magnitude smaller than the Larmor period at a magnetic field of $\Bext = 9\,$T. 
Hence, we can describe the action of a single pulse as an instantaneous mapping of the spin components before~($\S_\mathrm{b}$,~$\J_\mathrm{b}$) and after~($\S_\mathrm{a}$,~$\J_\mathrm{a}$) the pulse~\cite{yugov09,jasch17}
\begin{subequations}
	\begin{align}
	S^z_\mathrm{a} &= \frac{1}{4} + \frac{1}{2} S^z_\mathrm{b} \,,  &&\qquad S^x_\mathrm{a} = S^y_\mathrm{a} = 0 \,, \label{eq:pulse_S}\\	
	J^z_\mathrm{a} &= S^z_\mathrm{b} - S^z_\mathrm{a}\,, &&\qquad J^x_\mathrm{a} = J^y_\mathrm{a} = 0 \,. \label{eq:pulse_J}
	\end{align}
	\label{eq:pulse}%
\end{subequations}
In order to mimic the quantum properties of the spins, we must consider each pulse as a quantum mechanical measurement. 
According to the uncertainty principle, the semiclassical description of the pulses becomes nondeterministic. 
We model this uncertainty for $\S$ by normal distributions with mean values given by Eq.~\eqref{eq:pulse_S} and appropriate variance such that quantum mechanical property $\langle (\hat S^\alpha)^2\rangle = 1/4$, $\alpha \in \{x,y,z\}$, holds, see Ref.~\cite{scher20} for details. 
The pulse relations for the trion pseudospin~\eqref{eq:pulse_J} remain unchanged.

\subsection{Spin dynamics, spin mode locking, and nuclei-induced frequency focusing}
\label{sec:results_NIFF}

The time evolution of the spin dynamics is given by the ensemble average over ${M = 4800}$ independent classical trajectories with random initial conditions.
For this purpose, we use $2400-4800$ CPU cores on Hazel Hen, parallelized using pure MPI, i.\,e., we calculate $1$ or $2$ trajectories per core.

\begin{figure}[b!]
	\centering
	\subfloat{\includegraphics[width=0.46\columnwidth]{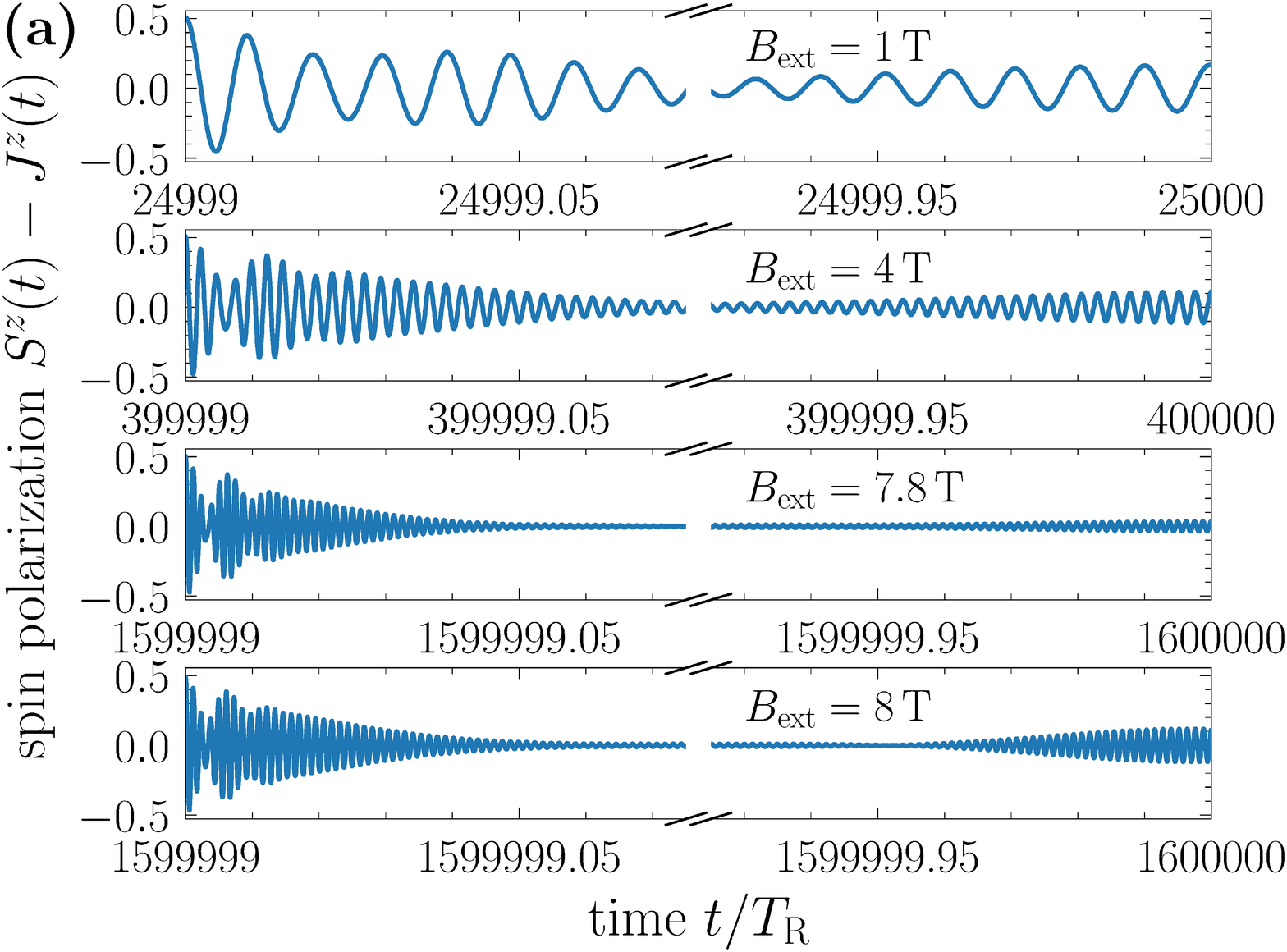} 	\label{fig:spindynamics}}	
	\subfloat{\includegraphics[width=0.54\columnwidth]{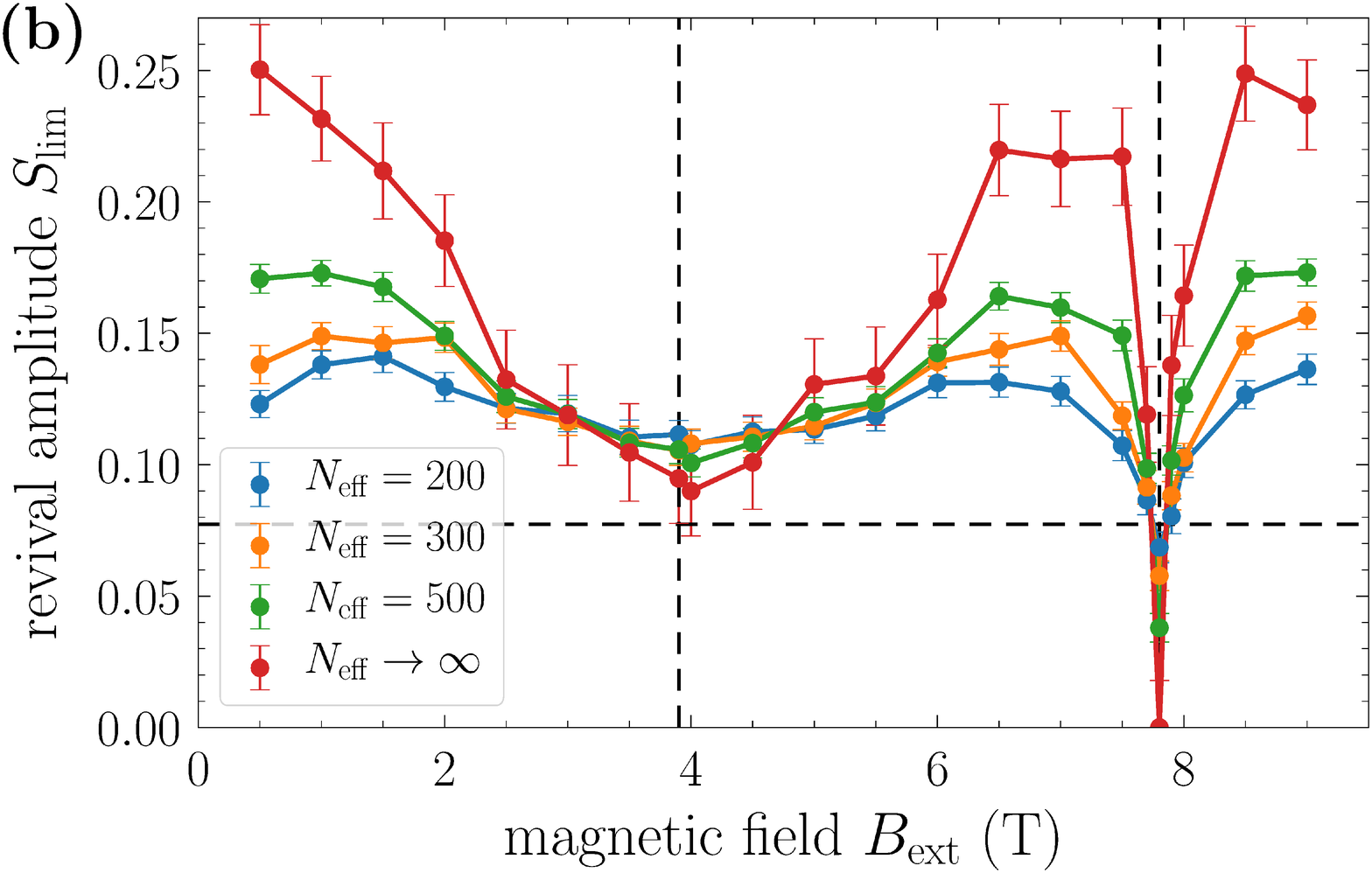} 	\label{fig:Slim_Bext}}	
	\caption{(a) Spin dynamics after a long train of periodic pulses with repetition period $\TR$ for various magnetic fields 
		$\Bext$ for an effective bath of $\Neff = 500$ nuclear spins. The oscillating spin polarization initially dephases, but a revival appears due to SML in combination with NIFF before the arrival of the next pulse. The amplitude of the revival signal does not change significantly anymore upon application of further pulses, i.\,e., the system is in a NESS.
		(b) Limiting values of the revival amplitude $\Slim$ as a function of the magnetic field $\Bext$ for various effective bath sizes 
		$\Neff$. The horizontal dashed line represents the revival amplitude which emerges even without NIFF.
		The vertical dashed lines represent values of $\Bext$ fulfilling the the nuclear resonance condition~\eqref{eq:NRC}. Further parameters: $\gh=0.66$, $\Delta \ge = 0.005$, $\Delta \gh = 0.016$. Similar figures are shown in Ref.~\cite{scher20}, but for different parameters.
	}\label{fig:S_NIFF}
\end{figure}

Examples for the spin dynamics for different magnetic fields after a long train of periodic pulses are shown in Fig.~\ref{fig:spindynamics}.
The initially created spin polarization precesses around the transverse magnetic field while
dephasing occurs on a timescale of $1\,$ns due to the fluctuations of the Overhauser field and due to the $g$ factor spread. 
The dephasing is faster for larger magnetic fields since a $g$ factor spread $\Delta g$ implies a term proportional 
to $\Delta g \Bext$ in the dephasing rate~\cite{greil06a}.
Modulations of the signal are discernible in the time-dependence of the combined quantity~${S^z - J^z}$, which is proportional to the signal measured in experiments~\cite{yugov09}. 
The modulations stem from the different Larmor frequencies of the electronic spin $\S$ and trion pseudospin~$\J$.

After the initial dephasing, a revival appears before the 
arrival of each next pulse due to spin mode locking~(SML).
An initial buildup of this revival happens already within $\mathcal{O}(10)$~pulses.
After a long train of pulses, the SML can change due to nuclei-induced frequency focusing~(NIFF).
Its amplitude depends on the strength of the external magnetic field due to its influence on NIFF, see Fig.~\ref{fig:S_NIFF}.
Note that for the data shown in Fig.~\ref{fig:spindynamics}, the revival amplitude does not change noticeably anymore upon further application of pulses because the NESS is already reached. To be precise, a quasistationary state is reached
which becomes apparent when studying the system stroboscopically at specific instants of the time interval between consecutive pulses, e.\,g., before the arrival of a pulse to study the revival amplitude.

\begin{figure}[b]
	\sidecaption
	\includegraphics[width=7.5cm]{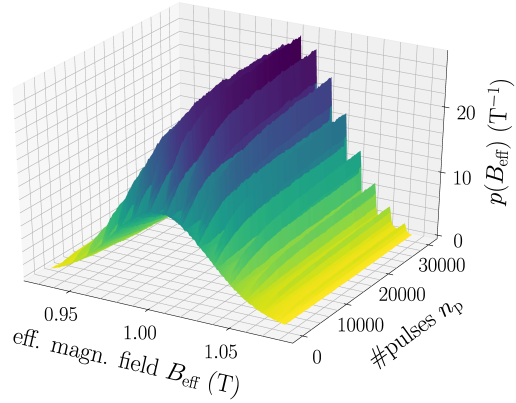}
	\caption{Nuclei-induced frequency focusing: Probability distribution of the effective magnetic field~$p(\Beff)$ as a function of the number of applied pulses $n_\mathrm{p}$. Initially ($\np = 0$), the effective magnetic field follows a simple normal distribution. Equidistant peaks emerge in the distribution due to the application of periodic pulses. The peak positions correspond to the values of $\Beff$ which fulfill the even resonance condition~\eqref{eq:ERC}. Parameters: $\Bext = 1\,$T, $\Neff=500$, $\gh=0.66$, $\Delta \ge = 0.005$, $\Delta \gh = 0.016$.}	
	\label{fig:Overhauser}
\end{figure}

This saturated revival amplitude $\Slim$ as a function of the magnetic field shows an interesting nonmonotonic dependence on the magnetic field which is depicted in Fig.~\ref{fig:Slim_Bext}.
Two pronounced minima are visible at positions corresponding to the nuclear resonance condition~\cite{beuge17,klein18,scher20}
\begin{equation}
\gn \mun \Bext \TR = \pi k\,, \qquad k \in \mathds{Z} \,.
\label{eq:NRC}
\end{equation}
The values of $\Bext$ fulfilling this condition are highlighted as vertical dashed lines in Fig.~\ref{fig:Slim_Bext}. 
The condition describes the number of half-turn revolutions of the nuclear spins in the external magnetic field $\Bext$ between consecutive pulses. 
Note that we consider only a single type of isotope for the nuclear spins here. 
Since generically there are several present in GaAs QDs, they can be viewed as an average isotope.
This simplification is lifted later in Sect.~\ref{sec:isotopes}.
The horizontal dashed line represents the SML revival amplitude which emerges already without any NIFF.
The comparison between the cases with and without NIFF is an important physical question addressed in Ref.~\cite{scher20}. 

The application of long trains of pulses leads to the emergence of a comblike structure in the probability distribution $p(\Beff)$ of the effective magnetic field.
This frequency focusing in the nuclear spin bath is what causes the nuclei-induced frequency focusing in the spin dynamics.
The buildup of the comblike structure is illustrated in Fig.~\ref{fig:Overhauser}.
The peak positions are found at the values of the effective magnetic field~$\Beff$ which fulfill the resonance condition
\begin{equation}
g_\mathrm{e}\muB\Beff \TR = 2 \pi k \,,  \qquad k \in \mathds{Z}\,. \label{eq:ERC}
\end{equation}
This condition describes full-turn Larmor periods of the electronic spin between consecutive pulses.
We refer to it as the \emph{even} resonance condition because $2k$ is an even integer.
In some cases, e.\,g., for $\Bext \approx 7.8\,$T, we find peaks at values of the effective magnetic field which fulfill the so called \emph{odd} resonance condition 
\begin{align}
g_\mathrm{e}\muB\Beff \TR = (2 k + 1)\pi \,, \qquad k \in \mathds{Z}\,. \label{eq:ORC}
\end{align}
It describes half-turn Larmor periods between consecutive pulses and leads to a reduced revival amplitude in comparison to the case without nuclei-induced frequency focusing, see Fig.~\ref{fig:S_NIFF} for $\Bext = 7.8\,$T.
Further insight into the importance of the two different resonance conditions is presented in 
Refs.~\cite{jasch17,beuge17,klein18,scher20}.

Furthermore, we find that the probability distribution of the effective magnetic field 
can shift as a whole; this effect is known as dynamic nuclear polarization~(DNP)~\cite{scher20}.
This polarization can be larger than the typical fluctuations of the Overhauser field, leading to a certain increase of the coherence time.
Reaching the corresponding NESS in the simulations requires two orders of magnitude more pulses 
than reaching steady values of the revival amplitude.
This renders the reliable simulation of the DNP unfeasible for magnetic fields much larger than $2\,$T
and imposes an important challenge for further improvements of the employed algorithms.

Generally, simulating large effective bath sizes and large magnetic fields is a tremendous challenge.
The number of pulses required to reach the steady values of the revival amplitude scales with $\Bext^2$ and linearly with $\Neff$~\cite{scher18,scher20}.
Moreover, in order to track the fast Larmor precession the integration step size decreases approximately with $\Bext^{-1}$ so that we are facing a cubic scaling in the computational complexity for larger fields. 
Developing performant approaches to mitigate this problem is part of our current research. 
A first approach of this kind, which is already applied in the present simulations, is briefly discussed in Sect.~\ref{sec:efficient}.

\subsection{Role of the nuclear spin bath composition}
\label{sec:isotopes}

Real QD ensembles studied in experiments do not consist of a single isotope but of many, e.\,g., 
they are GaAs or InGaAs QDs.
This can be accounted for in the equations of motion~\eqref{eq:eom}, but it increases the dimension of the 
ordinary differential equation system and becomes intractable when studying the exponential 
parameterization~\eqref{eq:couplings} of the hyperfine couplings.
A common simplification is the so called `box model', where all couplings are chosen to be equal~\cite{merku02}. 
Then, the full dynamics of the Overhauser field can be described by a single equation of motion for the subfield
made up of each different isotope in the system. 
For brevity, we omit the precise equations here.

We apply a simpler pulse model than in the previous Sect.~\ref{sec:model_NIFF} based on Ref.~\cite{scher18}, 
where we completely neglect the excitation of the trion state and instead describe the pulse action by the simple relation
\begin{equation}
S^z_\mathrm{a} = \frac{1}{2}\,, \qquad S^x_\mathrm{a} = X\,, \qquad S^y_\mathrm{a} = Y\,,
\end{equation}
with $X$ and $Y$ being random numbers sampled from a normal distribution around zero with variance $1/4$, i.\,e., 
we still consider each pulse as a quantum mechanical measurement.
We point out that this pulse model does not show SML without NIFF, i.\,e., for a small number of pulses no revival amplitude
prior to the next pulse occurs. 
Only long trains of pulses leading to NIFF engender a revival amplitude.
This is a fundamental difference to the more elaborate pulse model used in Sect.~\ref{sec:model_NIFF} 
and the main downside of neglecting the generation of spin polarization by means of an intermediate trion state.
Still, the nondeterministic description of the pulse is essential to mimic the 
quantum mechanical behavior~\cite{scher18,klein18,scher20}.
In contrast to the previous Sect.~\ref{sec:model_NIFF}, we omit the $g$ factor spread here, 
but it barely affects the NIFF behavior~\cite{scher20}.

\begin{figure}[t]
	\sidecaption[t]
	\includegraphics[width=7.5cm]{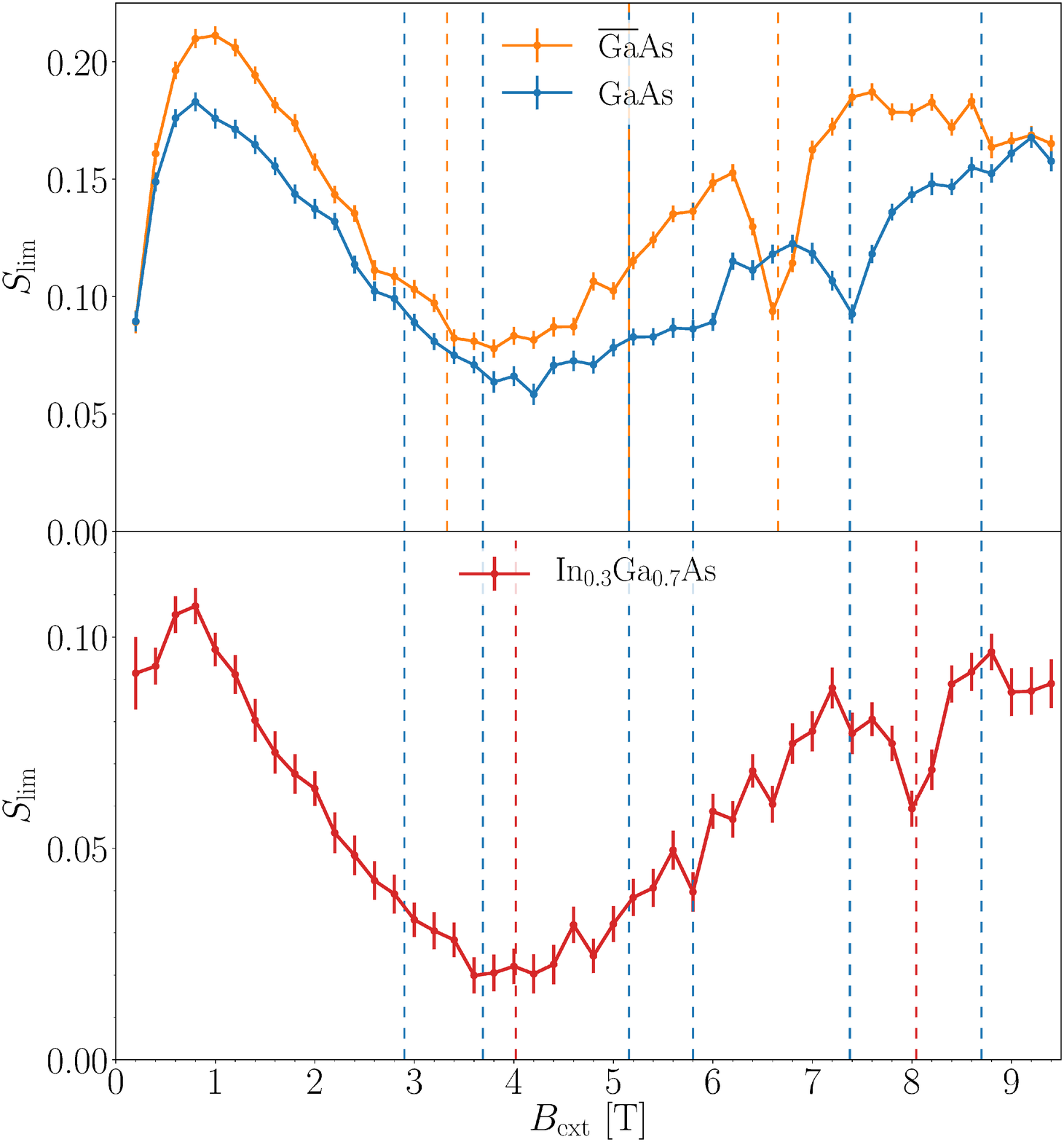}
	\caption{Saturated revival amplitude $\Slim$ as a function of the magnetic field $\Bext$ for different 
		compositions of the nuclear spin bath (\mbox{$N = 100$}). In the upper panel the dephasing time is 
		$\Tnstar = \sqrt{2}\,$ns, in the lower panel it is $\Tnstar = 1\,$ns. The horizontal dashed lines 
		indicate the values of $\Bext$ which fulfill the nuclear resonance condition~\eqref{eq:NRC} for the various isotopes.}	
	\label{fig:isotopes}
\end{figure}

When studying nuclear spin baths consisting of various isotopes, 
we expect additional nuclear resonance conditions~\eqref{eq:NRC} to play a role 
because each isotope has a different gyromagnetic ratio $\gn \mun$.
Figure~\ref{fig:isotopes} shows the magnetic field dependence of the revival amplitude for three 
different compositions of the nuclear spin bath.
In the upper panel, we compare the result for a GaAs QD with the case where a GaAs QD with an average Ga isotope is considered ($\overline{\text{Ga}}$As). In the lower panel, we show 
the result for an In$_{0.3}$Ga$_{0.7}$As QD. 
A quantitative comparison of both panels is not possible yet because the dephasing times $\Tnstar$, 
which determine the width of the Overhauser field, are chosen differently.
The main conclusion is that not all nuclear resonance conditions play a major role as we still mainly find 
a broad minimum around $\Bext=4\,$T and a sharper minimum at larger magnetic fields.
Some additional structure is visible around $\Bext = 5.8\,$T, but further more accurate calculations 
are required for a better resolution.
Due to the increased complexity of the physical situation, the degree of NIFF is 
reduced when the number of different isotopes is increased.
Furthermore, indium dominates the behavior of the system even for very small concentrations 
because it has spin $9/2$ whereas the other isotopes have spin $3/2$, and its hyperfine coupling constant 
is also larger~\cite{coish09}.

\section{Spin inertia and polarization recovery in quantum dots}
\label{sec:RSA}

In the previous sections, we considered the application of a transverse magnetic field (Voigt geometry).
In a different class of pump-probe experiments on QD ensembles, a longitudinal magnetic field 
(Faraday geometry) is applied.
Here, we consider QDs charged by either localized electrons or holes.
Typical experiments study the occuring spin inertia and polarization 
recovery effects~\cite{heisterkamp15,zhukov18,smirnov18,scher19}. 
Periodic circularly-polarized laser pulses are applied with repetition period $\TR=13.2\,$ns, but their 
helicity is modulated between $\sigma^+$ and $\sigma^-$ with frequency~$\fm$.
Studying the spin polarization as a function of this modulation frequency shows the so called spin inertia effect: 
when the modulation frequency is increased, the spin polarization decreases. 
This can be understood as an inertia of the spin which prevents it from following 
the switching of the pulse helicity arbitrarily quickly.
This effect enables the measurement of slow relaxation times $\mathcal{O}(\mu\text{s})$ of the system.
The polarization recovery effect consists of the increase of the spin polarization upon an increase the
longitudinal magnetic field. 
Hence, the polarization recovery curve is the graph of the spin polarization as a function of this magnetic field.
We report on the influence of the pumping strength on these experiments; for details see Ref.~\cite{scher19}.

The simulation of this setup is less demanding than the one of Sect.~\ref{sec:NIFF} because the 
dynamics of the Overhauser field plays a minor role so that it can be 
considered as frozen~\cite{merku02}, i.\,e., as static, but still random according to a normal distribution to account for its statistical fluctuations.
Then, the spin dynamics of the localized charge carrier in the ground state of each QD is described by
\begin{equation}
\label{eq:dSdt}
\ddt \S = \left(\bm\Omega_\mathrm{N,g}+\bm\Omega_\mathrm{L,g}\right)\times\S 
- \frac{\S}{\tausg} + \frac{J^z}{\tau_0} \ez \,,
\end{equation}
where $\bm\Omega_\mathrm{N,g}$ is the frequency of the spin precession caused by the Overhauser field,
$\bm\Omega_\mathrm{L,g}= \Omega_\mathrm{L,g}\ez = g_\mathrm{g} \mu_\mathrm{B} \Bext \ez$ 
is the Larmor frequency, with $g_\mathrm{g}$ being the effective longitudinal $g$ factor of the ground state, and $\Bext\ez$ the external longitudinal magnetic field.
Furthermore, the phenomenological term $-\S/\tausg$ describes the spin relaxation unrelated to the hyperfine interaction with the nuclear spins in the QD. 
The dynamics of the trion pseudospin between the pump pulses is described similarly to Eq.~\eqref{eq:dSdt} by the equation of motion
\begin{equation}
\label{eq:dJdt}
\ddt \J =\left(\bm\Omega_\mathrm{N,t}+\bm\Omega_\mathrm{L,t}\right)\times \J - \frac{\J}{\taust} - \frac{\J}{\tau_0} \,.
\end{equation}
The Overhauser field is again normal distributed, but it can be anisotropic for hole spins~\cite{testelin07}, and the spin polarization of the QD ensemble is calculated by averaging over all trajectories stemming from random initial conditions. 
Details of the simulation for QDs charged by either electrons or holes can be found in the original publication~\cite{scher19}.
The computations are not as expensive as in the previous sections because we only need to study magnetic fields up to $300\,$mT and overall a smaller number of pulses.
The periodic laser pulses are described by a similar but more general relation than given by 
Eq.~\eqref{eq:pulse}, which includes possible rotations of the transverse spin components 
and also allows us to consider various pump pulse efficiencies~\cite{scher19}.

\begin{figure}[t!]
	\centering
	\subfloat{\includegraphics[width=0.48\columnwidth]{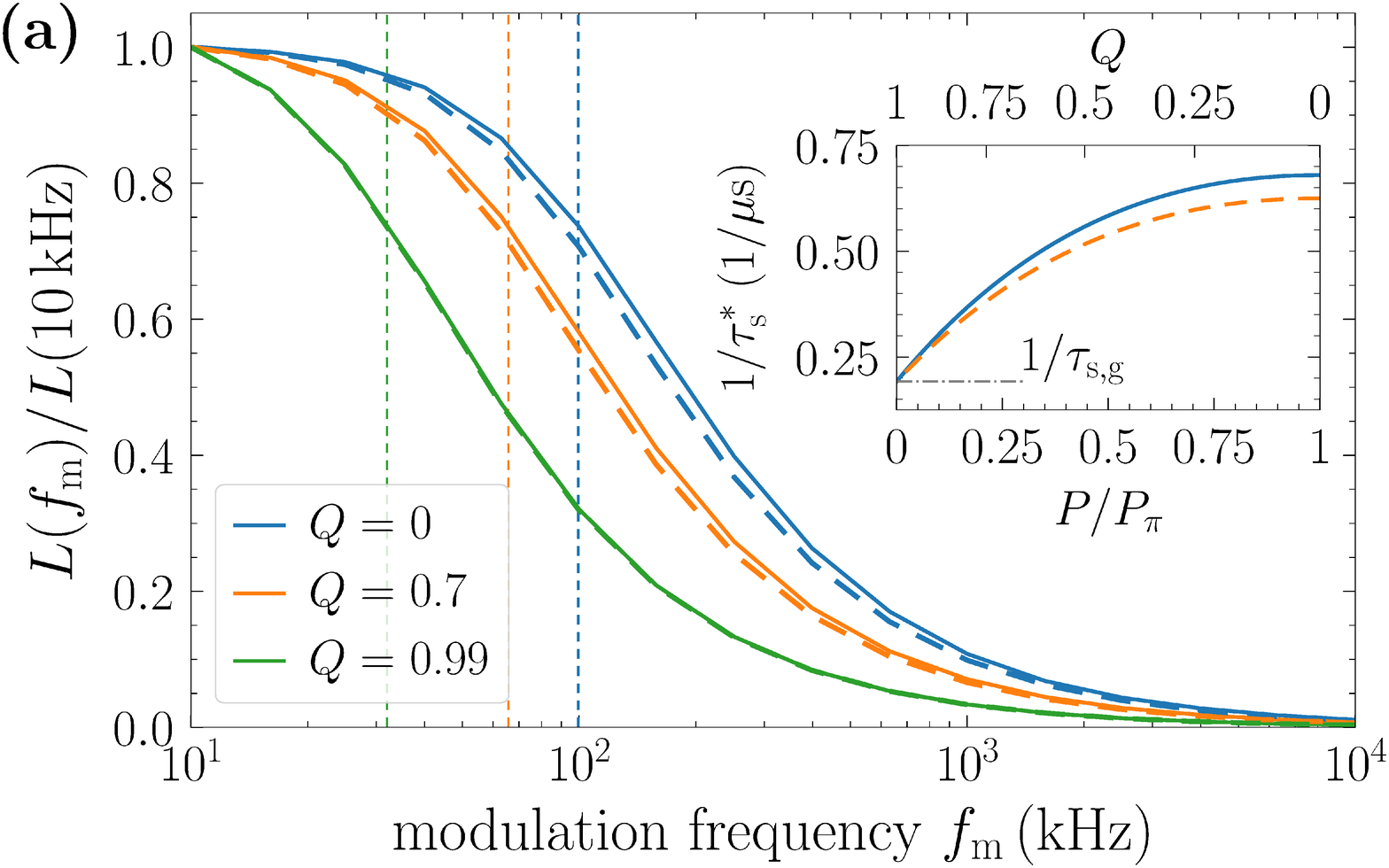} 	\label{fig:SpinInertia}}
	\subfloat{\includegraphics[width=0.48\columnwidth]{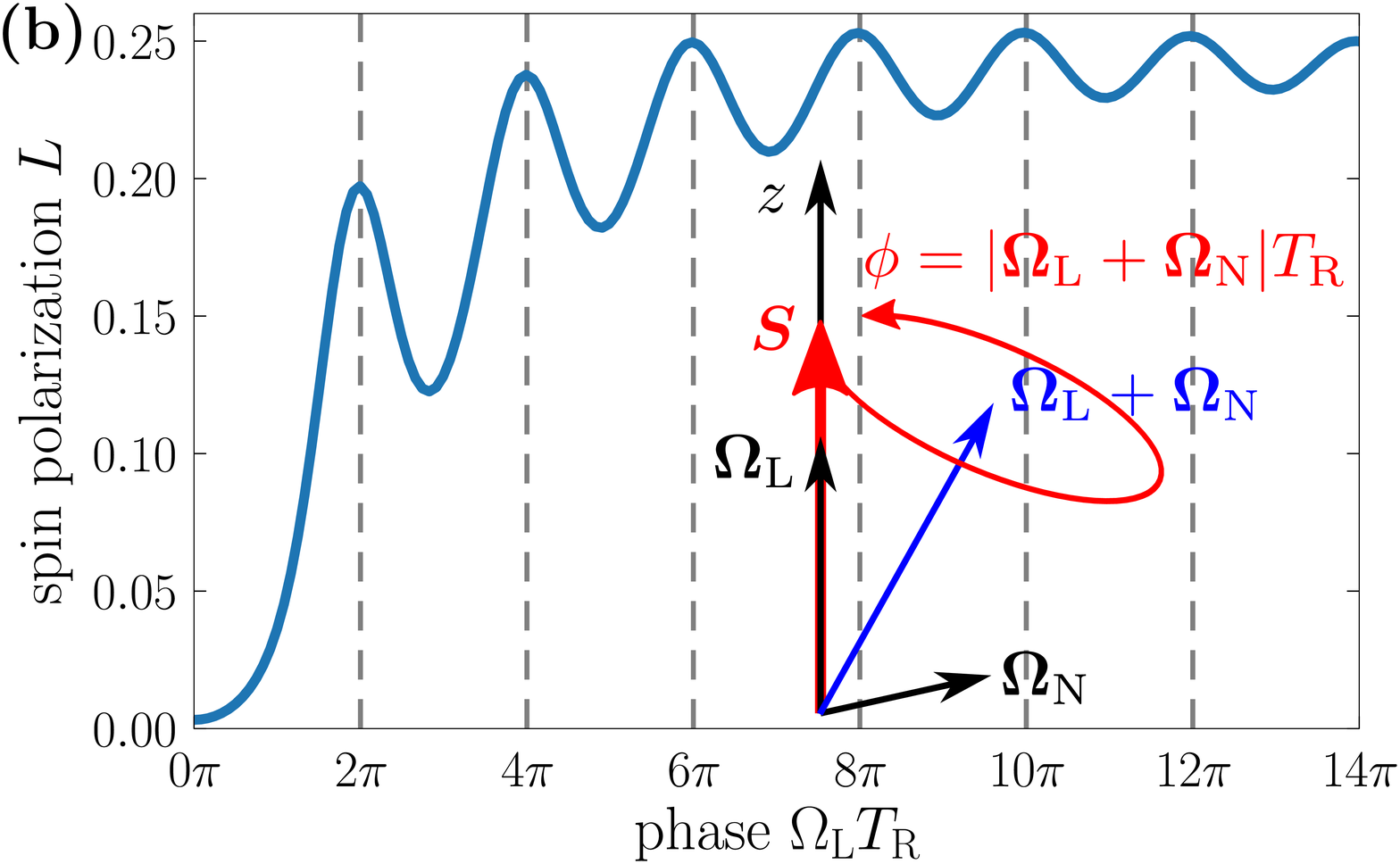} 	\label{fig:RSA}}
	\caption{(a)~Spin inertia effect for localized holes in quantum dots: spin polarization $L$ as a function of the modulation frequency $\fm$ for a magnetic field of $\Bext = 300\,$mT and various pumping efficiencies $Q$. The vertical dashed lines represent the typical cut-off frequencies~$1/(2\pi \tauseff)$. The inset shows the inverse of the effective spin relaxation time $\tauseff$ as a function of the pump power $P$ and of the pump efficiency $Q$. The dashed curves are calculated analytically for the limit $\Bext \to \infty$. (b)~Illustration of resonant spin amplification in Faraday geometry: the spin polarization increases as a function of the phase $\Omega_\mathrm{L} \TR \equiv \Omega_\mathrm{L,g} \TR \propto \Bext$ with periodic oscillations at positions fulfilling the resonance condition~\eqref{eq:RSA}. Both figures are taken from Ref.~\cite{scher19}, licensed under \href{https://creativecommons.org/licenses/by/4.0/}{CC BY 4.0}, with slight modifications to the layout. \label{fig:SpinInertia_RSA}}
\end{figure}

Our main results from Ref.~\cite{scher19} are the following.
In accordance with experiments~\cite{zhukov18}, we find that applying a larger pump power leads to a decrease of the effective spin relaxation time in the system.
This behavior is visualized in the inset of Fig.~\ref{fig:SpinInertia} for the case of localized holes in the QDs. The inverse of the effective spin relaxation time is plotted as a function of the pump power~$P$ and of the pumping efficiency~$Q$ for a magnetic field of $\Bext = 300\,$mT. The best pumping efficiency is achieved for $Q=0$, which describes so called $\pi$~pulses reached at the pump power~$P_\pi$. Weak pulses are described by the limit $Q \to 1$.
A linear extrapolation to zero pump power yields the equilibrium spin relaxation time~$\tausg$.
The main panel of Fig.~\ref{fig:SpinInertia} shows the spin inertia effect for various pumping efficiencies~$Q$.
Upon increasing the modulation frequency $\fm$, the spin polarization decreases. 
Analytical results for the limit $\Bext \to \infty$ (dashed curves) support our findings. 
The effective spin relaxation time can be extracted from the dependence of the spin polarization on the modulation frequency.

Furthermore, we analyze the role of the saturation of spin polarization in the polarization recovery measurements.
We find that approaching the saturation limit of the spin polarization leads to a broadening of the typical V-like shape of the polarization recovery curves (the curves are symmetric with respect to the magnetic field $\Bext$), similar to what is observed experimentally~\cite{zhukov18}.

Most importantly, we find the emergence of resonant spin amplification in Faraday geometry~\cite{scher19}.
It is a well established effect in Voigt geometry~\cite{kikkawa98,yugov12} and also known for a tilted magnetic field~\cite{zhukov12}, but it is to our knowledge not yet discovered in a pure longitudinal field configuration.
It can emerge purely due to the transverse fluctuations of the Overhauser field 
in the QDs under certain conditions and 
it can be exploited to measure the longitudinal~$g$~factor~$g_\mathrm{g}$ of the localized charge carriers in the QDs.
We stress, however, that the longitudinal fluctuations of the Overhauser field may not be too large
since the effect is smeared out otherwise.
A typical polarization recovery curve illustrating this effect is shown in Fig.~\ref{fig:RSA}. 
The spin polarization increases for larger magnetic fields $\Bext$, which is proportional to the phase ${\Omega_\mathrm{L} \TR \equiv \Omega_\mathrm{L,g} \TR} \propto \Bext$ displayed on the abscissa.
Periodic oscillations are found whenever the resonance condition
\begin{equation}
\Omega_\mathrm{L,g} \TR = 2\pi k \,, \qquad k \in \mathds{Z} \,, \label{eq:RSA}
\end{equation}
is fulfilled. 
This effect is caused by the slight tilt of the effective magnetic field $\bm\Omega_\mathrm{L,g} + \bm\Omega_\mathrm{N,g}$ from the $z$ axis due to the transverse components of the Overhauser field $\bm\Omega_\mathrm{N,g} \equiv \bm\Omega_\mathrm{N}$ in each QD, see the sketch in Fig.~\ref{fig:RSA} for a graphical illustration.
We estimate that resonant spin amplification in Faraday geometry can be observed under the 
condition $\wn \lesssim \sqrt{2} \pi/\TR$, where $\wn$ is the typical fluctuation strength of the Overhauser field, 
when sufficiently strong pump pulses are applied. 
This condition implies that reducing the pulse repetition period $\TR$ can help to reveal this new effect.
In the experiments of Ref.~\cite{zhukov18}, the effect is not observed because the condition is not fulfilled and pump pulses of low power are used.

Note the similarity to the resonance conditions discussed in Sect.~\ref{sec:NIFF}. 
The resonance conditions are central to understanding the nonequilibrium spin dynamics in QDs 
subjected to trains of periodic pulses whenever an external magnetic field is applied.
Preliminary experimental results indicate that the effect depicted in Fig.~\ref{fig:RSA}
can indeed be measured.

\section{Efficient simulations}
\label{sec:efficient}

Solving ordinary differential equations~(ODEs) is straightforward, especially if no stability problems occur as in our case.
We apply the Dormand-Prince method as ODE solver, which is an adaptive fifth-order Runge-Kutta algorithm, 
using the implementation provided in Ref.~\cite{NumericalRecipes}.
The simulation of realistic experimental setups is extremely challenging because the relevant time scales 
govern several orders of magnitude.
For this reason, on the one hand, the integration error must be small enough so that errors do not add up significantly 
after up to millions of pulses.  
On the other hand, one cannot aim at reducing it too much by very small time steps because this would spoil the performance.

Reaching a NESS as required in Sect.~\ref{sec:NIFF} for large magnetic fields to study the NIFF behavior is extremely demanding because the computational complexity scales with~$\Bext^3$. 
Simulating large spin baths is an additional challenge due to a linear scaling with $\Neff$ of the numbers 
of pulses required to reach the quasistationary states.
In Ref.~\cite{scher19}, we also established scaling laws which allow us to 
extrapolate to infinite bath sizes $\Neff \to \infty$.
The direct simulation of such bath sizes is not possible even by high performance computing.

\subsection{Spectral density and rotating frame approach}

As discussed in Sect.~\ref{sec:model_NIFF}, the application of an efficient approach~\cite{fauseweh17} 
to the ODE system~\eqref{eq:eom} reduces its dimension from $3N+6$ to $3\Ntr+6$, with the truncation 
parameter $\Ntr \ll N$, by replacing sums of bath spins by auxiliary vectors.
In a typical scenario, $\Ntr = 75$ for an effective bath size of $\Neff = 200$ with $N \to \infty$ is used. 
If we solved the normal ODE system~\eqref{eq:eom} for $N=200$ nuclear spins, the dimension would be about $2.5$ times larger, eventually requiring the usage of the slower L2 and L3 caches for large spin baths.

Further performance improvements are gained by solving the ODE system~\eqref{eq:eom} in a rotated frame.
The electronic spin $\S$ and the trion pseudospin~$\J$ mainly precesses around the large transverse external magnetic field, and this precession is by far the fastest frequency in the system. 
Thus, it determines the integration step size in a linear manner, i.\,e., it is proportional to $\Bext^{-1}$.
However, this precession motion can be easily described analytically. 
The application of such an ansatz to the ODE system yields modified equations of motion, and the integration 
of the ODE system is about three times faster while maintaining the same numerical accuracy.
Unfortunately, the step size still decreases linearly when increasing the magnetic field strength:
the nuclear spins also show the fast oscillations due to their coupling to the precessing electronic spin, 
which need to be resolved numerically.

Establishing even more efficient approaches is part of our current research.
One particular goal is an approach in which the integration step size is not determined 
linearly by the magnetic field strength since this would render the simulation of 
large magnetic fields much more efficient.

\subsection{Single thread performance}

Vectorization is the key to achieve good single thread performance.
The applied Runge-Kutta algorithm is easily vectorized by any modern compiler.
For the ODE system~\eqref{eq:eom}, a long loop must iterate over all nuclear spins, but the structure of their equations of motion is identical.
By using a structure of arrays for the data layout, good vectorization is accomplished.
Due to the rather small dimension of the ODE system (about 230 equations), the calculations are operating on the L1~cache. 
We measure a double performance of 11.2~GFLOPS per core on a dual socket Intel~Xeon~Broadwell~E5-2680~v4~(2x14 cores, 2.4~GHz) without hyperthreading, with a vectorization ratio of 91\%.

Vectorization is not straight forward when we apply the box model to the nuclear spins as in Sect.~\ref{sec:isotopes} because there is no long loop which iterates over the nuclear spins. 
The typical dimension of a single ODE system of 9 to 18 is small.
But we have to solve the ODE system for $\mathcal{O}(10^4)$ independent initial conditions and thus, we can group them together and solve their equations of motion simultaneously, i.\,e., we consider the group as a single combined ODE system. 
Again, using a structure of arrays for the data layout, good vectorization of the code is accomplished. 
The maximal group size is limited by the size of the L1~cache; the precise group size is chosen to optimize the performance.

The same principle is applied when solving the ODE sytem of Sect.~\ref{sec:RSA} with a dimension of 6.
When building groups of 64 independent random initial configurations, we measure a double precision performance of 9.7~GFLOPS per core on a dual socket Intel~Xeon~Broadwell~E5-2680~v4~(2x14 cores, 2.4~GHz) without hyperthreading, with a vectorization ratio of 99\%.
Without the grouping procedure, the performance decreases to mere 2.0~GFLOPS per core with a vectorization ratio of only 13\%.

\subsection{Parallelization}

The semiclassical simulation of the spin dynamics is easily parallelizable using pure MPI since the calculation of a single trajectory does not depend on the other ones.
Minor communication takes place only at the end of the simulation where ensemble averages are calculated and the Overhauser fields are stored for the subsequent statistical analysis. 
The influence of the minor I/O on the performance is negligible.

\begin{figure}[h]
	\sidecaption
	\includegraphics[width=7.5cm]{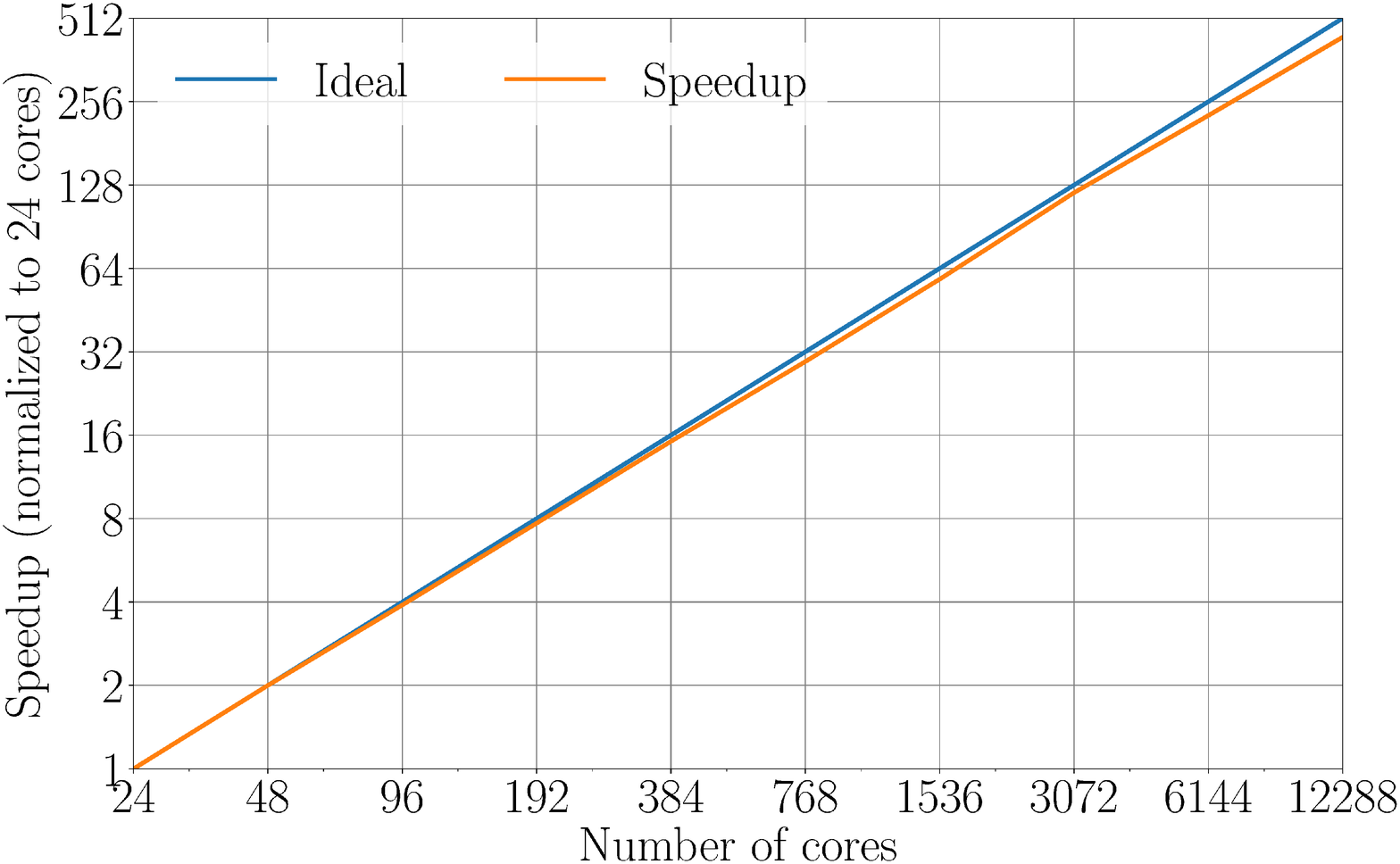}
	\caption{Logarithmic plot of the scaling behavior of our code used in Sect.~\ref{sec:results_NIFF} 
		on Hazel Hen (2x~Intel~Xeon~Haswell~E5-2680~v3~(2.5~GHz, 12 cores) per node) at HLRS. The relative 
		speedup normalized to 24 cores is shown and compared to the ideal case. In this benchmark, a total of $M=24576$ trajectories are calculated.} \label{fig:scaling}
\end{figure}

Figure~\ref{fig:scaling} shows a typical scaling behavior of our simulation code used for the simulations presented in Sect.~\ref{sec:results_NIFF}. 
Great scaling is achieved up to 12288 cores when calculating $M=24576$ trajectories.
In this extreme scenario, two trajectories are calculated per core. 
Due to the adaptive integration of the ODE system, some integrations finish sooner than others, which leads to a 
slightly reduced parallel efficiency when only few trajectories are calculated per core.
In practice, we typically use $2400$ cores to calculate $4800$ independent trajectories, i.\,e., we calculate two trajectories per core.
For large magnetic fields combined with large bath sizes, it is required to use $4800$ cores for the same number of trajectories such that the simulation finishes within~$24\,$h.

In cases where a deterministic pulse model is applied, a slightly better load balance can be achieved by running a short benchmark of about ten pulses for the calculation of each trajectory in the beginning, i.\,e., before the actual simulation. Then, slow and fast calculations can be grouped together to improve the load balance.
This procedure does not work reliably when the pulses are nondeterministic because the short benchmarks are not representative anymore for the full integration.
Generally, when more than one trajectory is calculated per core, the load balance improves automatically due to a self-averaging effect.

As demonstrated in Fig.~\ref{fig:scaling}, the number of total cores can be increased well beyond the $4800$ cores typically used in our simulations if a better statistical accuracy, accomplished by a larger ensemble size $M$, is desired. The statistical errors of the ensemble averages are proportional to $1/\sqrt{M}$.

\section{Conclusion}

Our large-scale simulations of the nonequilibrium spin dynamics in quantum dots~(QDs) subjected to trains of periodic laser pulses pave the way for a better understanding and description of related pump-probe experiments~\cite{greil06a,greil06b,greil07a,varwig14,zhukov18,klein18}.
By combining sophisticated efficient approaches with the raw computation power of Hazel Hen provided by the HLRS, we are able to reach nonequilibrium (quasi)stationary states for the full range of magnetic fields studied in the experiments for large bath sizes.
Further scaling relations are established which enable us to extrapolate to infinite bath sizes, which is the physical limit of interest.

Meanwhile, we explore the influence of the nuclear spin bath composition and find that for InGaAs QDs, indium plays the most important role due to its large spin $9/2$ and its larger hyperfine coupling.
The degree of nuclei-induced frequency focusing is reduced when more complex compositions are considered.
The combination of this enhanced and more realistic description of the nuclear spin bath 
with our more elaborate model of Sect.~\ref{sec:model_NIFF} comprising the generation of spin polarization via trion excitation is an imminent part of our current research.

The direct numerical simulation of the spin inertia and polarization recovery experiments for 
arbitrary pumping strength yields a better understanding of the experimental results presented in Ref.~\cite{zhukov18}. 
The existing analytic theoretical description~\cite{smirnov18} is only valid in the low pump power limit. 
We identify the influence of large pump powers leading to saturation effects in the measurements.
Importantly, we find the emergence of resonant spin amplification in Faraday geometry.
It allows for measuring the longitudinal $g$ factor of the localized charge carriers in the QDs.
Preliminary results from our experimental colleagues suggest that they can indeed detect the predicted effect.
The quantitative description of their measurements requires the extension of the current model to account for the inhomogeneous broadening of the trion transition energy~\cite{yugov09}, which is present in any real QD ensemble. 
This extension adds another stochastic component to the simulation, leading to a further increase of the 
computational complexity which we will address with the help of high performance computing.

\begin{acknowledgement}
We thank Carsten Nase and Dmitry S. Smirnov for helpful discussions and gratefully acknowledge the Gauss Centre for Supercomputing e.V. for supporting this project by providing computing time on the GCS Supercomputer Hazel Hen at H\"ochstleistungsrechenzentrum Stuttgart (HLRS). This study has been funded by the German Research Foundation (DFG) and the Russian Foundation for Basic Research (RFBR) in the International Collaborative Research Centre TRR~160 (Projects No.~A4 and No.~A7).
\end{acknowledgement}

\end{document}